\documentclass[12pt,amsfonts,amsmath,amssymb,color]{article}
\usepackage{authblk}
\def\baselinestretch{1.0}
\usepackage{amsmath,amssymb,amsfonts,latexsym,float,graphics,epsfig}
\usepackage{subfig}
\usepackage{verbatim}
\usepackage{relsize}
\usepackage{hyperref}
\usepackage{graphicx}
\usepackage{bm}
\usepackage{epstopdf}
\usepackage{slashed}
\usepackage{color}
\usepackage{tikz-cd}
\usepackage[utf8]{inputenc}
\usepackage[T1]{fontenc}

\begin{document}

\renewcommand\theequation{\arabic{section}.\arabic{equation}}
\catcode`@=11 \@addtoreset{equation}{section}

\newtheorem{axiom}{Definition}[section]
\newtheorem{theorem}{Theorem}[section]
\newtheorem{axiom2}{Example}[section]
\newtheorem{lem}{Lemma}[section]
\newtheorem{prop}{Proposition}[section]
\newtheorem{cor}{Corollary}[section]

\let\endtitlepage\relax

\begin{titlepage}
\begin{center}
\renewcommand{\baselinestretch}{1.5} 

{\Large \bf{Invariant measures for some dissipative}}\\
 {\Large \bf{  systems from the Jacobi last multiplier}}

\vspace{9mm}
\renewcommand{\baselinestretch}{1}  

\centerline{\large{\bf Aritra Ghosh}}

\vspace{5mm}
\normalsize
\text{School of Basic Sciences,}\\
\text{Indian Institute of Technology Bhubaneswar,}\\
\text{Jatni, Khurda, Odisha 752050, India}\\
\vspace{5mm}

\centerline{\large{ag34@iitbbs.ac.in; aritraghosh500@gmail.com}}

\begin{abstract}
Hamiltonian dynamics describing conservative systems naturally preserves the standard notion of phase-space volume, a result known as the Liouville's theorem which is central to the formulation of classical statistical mechanics. In this paper, we obtain explicit expressions for invariant phase-space measures for certain (generally dissipative) mechanical systems, namely, systems described by conformal vector fields on symplectic manifolds that are cotangent bundles, contact Hamiltonian systems, and systems of the Li\'enard class. The latter class of systems can be described by certain generalized conformal vector fields on the cotangent bundle of the configuration space. The computation of the invariant measures is achieved by making use of the formalism of Jacobi last multipliers.
\end{abstract}
\end{center}
\end{titlepage}

\section{Introduction}
It is known that standard Hamiltonian dynamics is formulated on phase spaces that are symplectic manifolds, i.e., even-dimensional smooth manifolds equipped with a non-degenerate and closed two-form $\omega$ \cite{arnold}. A powerful feature of this construction is the invariance of the volume-form $\omega^n = \omega^{\wedge n}$ under a Hamiltonian phase flow. This result which is called Liouville's theorem makes way for the formulation of classical statistical mechanics. The descriptions of dissipative systems are, however, quite different. For instance, some dissipative systems may be described by employing certain `generalized' Hamiltonian frameworks such as by resorting to conformal Hamiltonian dynamics \cite{conf,conf1} or by formulating Hamiltonian dynamics on contact or cosymplectic manifolds (see for example, \cite{CM1})\footnote{It may be remarked that conformal Hamiltonian dynamics and contact Hamiltonian dynamics can also be used to describe certain dynamical equations relevant in biology and pattern formation \cite{biology}.}. 

\vspace{2mm}

While some simple dissipative systems where a `linear' friction term appears with a constant coefficient can be described on symplectic manifolds using the notion of conformal Hamiltonian dynamics \cite{conf,conf1}, a more general framework for dissipative systems is provided by contact geometry \cite{Geiges} where one identifies the phase space of the system to be odd-dimensional and equipped with the so-called contact structure. In this setting, a suitably-adapted generalization of Hamiltonian dynamics may describe certain dissipative mechanical systems in a natural manner \cite{CM1,CM2,CM3,CM4,CM5}. The dynamics is also accompanied by the non-conservation of the volume-form if defined in a certain `standard' way (to be described later). However, it is possible to describe certain non-trivial invariant measures on phase spaces that are contact manifolds \cite{brav1,brav2} and this makes way for a corresponding formulation of statistical mechanics \cite{bravstat}.  

\vspace{2mm}

Apart from linearly-damped systems with constant damping strength, another class of systems where the second-order equation of motion breaks time-reversal invariance are systems of the Li\'enard class, being described by the equation\footnote{It may be noted that despite the presence of the damping term `$f(x) \dot{x}$', the dynamics may not always be dissipative. For example, the nonlinear Li\'enard system $\ddot{x} + kx \dot{x} + \omega^2 x + \frac{k^2}{9} x^3 = 0$ admits oscillatory (non-decaying) solutions \cite{chandra}.}
\begin{equation}\label{Lienard1definition}
\ddot{x} + f(x) \dot{x} + g(x) = 0,
\end{equation} where $f(x)$ and $g(x)$ are suitable (usually smooth) real-valued functions of the real variable $x$. Here, the damping strength may depend on the variable $x$, thereby giving rise to the possibility of nonlinear dynamics as well as that of limit cycles \cite{limit}. The phase-space flows admit a non-trivial divergence unlike conservative Hamiltonian systems respecting traditional Liouville's theorem. As may be expected, such systems cannot be described using the standard Hamiltonian approach although some such systems may be described by using nonstandard forms of Lagrangian and Hamiltonian functions \cite{CM3,CM5} provided $f(x)$ and $g(x)$ satisfy the so-called Chiellini integrability condition:
\begin{equation}\label{CC0}
\frac{d}{dx} \bigg( \frac{g(x)}{f(x)} \bigg) + l (l+1) f(x) = 0,
\end{equation} where $l$ is a real number.

\vspace{2mm}

The aim of this paper is to present expressions for invariant phase-space measures for some dissipative systems, focusing on three classes of systems -- (a) systems that are  described by conformal Hamiltonian dynamics, (b) systems that are described by contact Hamiltonian dynamics, and (c) systems of the Li\'enard class as defined above. For this purpose, we shall make use of the notion of Jacobi last multipliers \cite{whittaker}. In particular, we will re-derive the so-called `canonical' invariant measure for contact Hamiltonian systems \cite{brav1} by using the formalism of Jacobi last multipliers. Further, we shall describe a certain generalization of conformal Hamiltonian dynamics on `exact' symplectic manifolds (specifically, cotangent bundles) which will allow us to discuss Li\'enard-type systems. In particular, it will be shown that when $f(x)$ and $g(x)$ satisfy the Chiellini integrability condition (\ref{CC0}) (see for example, \cite{CM3}), it is possible to find analytical expressions for phase-space measures which are invariant under the corresponding dynamics. 

\vspace{2mm}

The paper is organized as follows. In the next section [Sec. (\ref{presec})], we will recall some basic definitions regarding Jacobi last multipliers and the Hamiltonian formalism on symplectic and contact manifolds. Following this, in Sec. (\ref{invariantsec1}), we shall present invariant phase-space measures for both conformal Hamiltonian dynamics as well as contact Hamiltonian dynamics. Finally, in Sec. (\ref{invariantsec2}), we will present a generalized version of conformal Hamiltonian dynamics which can describe Li\'enard-type systems [Eq. (\ref{Lienard1definition})] and will also present the explicit expressions for invariant phase-space measures when the Chiellini integrability condition (\ref{CC0}) is satisfied. We will conclude the paper with an extended discussion in Sec. (\ref{summary}). 

\section{Preliminaries}\label{presec}
In this section, we will review some well-known notions which will be useful for our analysis in the subsequent sections. This will also help us set the notation and clarify the basic conventions. We will begin by reviewing the framework of Jacobi last multipliers below which will be followed by a review of the Hamiltonian frameworks on symplectic and contact manifolds. 

\subsection{Jacobi last multiplier}
The last multiplier which was introduced by Jacobi in 1844 (see the classic text \cite{whittaker}) has turned out to be a useful tool in analytical mechanics. On one hand, the notion of Jacobi last multipliers allows one to deduce a conserved quantity (at least, locally) for a planar dynamical system or equivalently, for a dynamical system on an $m$-dimensional phase space if $(m-2)$ conserved quantities are already known -- in that case, the system can be reduced to a family of planar ones \cite{whittaker,Akash}. On the other hand, a last multiplier allows one to find the Lagrangians describing certain second-order differential equations \cite{Leach1,Leach2,Leach3,AGC,JLMLienard,mitra24} (see also, the older work \cite{yan78}), most notably, some from the Li\'enard class of systems \cite{JLMLienard,PG3,PG4}. More recently, it has found use in the context of non-holonomic Lagrangian mechanics \cite{hamel}, non-divergence-free vector fields \cite{carinenahojman}, and also in the context of Lie symmetries \cite{Nucci,JLMNew} and integrability \cite{carinenaintegrability}. 

\vspace{2mm}

Consider a dynamical system, i.e., a system of first-order equations which go as $\dot{x}_i = X_i(x_1,x_2, \cdots, x_m)$, where $i = 1,2,\cdots, m$. Here, $x_i$ could be thought of as being the (possibly local) coordinates in some region $U \subseteq \mathbb{R}^m$ and ${X_i}$ are real-valued and differentiable functions defined on $U$. The dynamical system is described by a vector field that goes as
\begin{equation}\label{Xvectorfield}
X = X_i \frac{\partial}{\partial x_i}, \quad \quad i = 1,2,\cdots,m. 
\end{equation}
Note that we have the definition $X(F) = \frac{dF}{dt}$, for a function $F$ defined on $U$. Let us consider the volume-form $\Omega = dx_1 \wedge dx_2 \wedge \cdots \wedge dx_m$. The divergence of $X$ can be defined from the Lie derivative as $\pounds_X \Omega = ({\rm div} X) \Omega$; if the divergence is zero, then the vector field is volume preserving. In general, ${\rm div} X \neq 0$. We can now define a Jacobi last multiplier (see \cite{hamel,carinenahojman} for discussions on the definition and results presented below). 

\begin{axiom}
A Jacobi last multiplier $M$ is a factor such that $MX$ has zero divergence. 
\end{axiom}

\begin{cor}
If $M \neq 0$ is a last multiplier of a dynamical system on an $m$-dimensional phase space as described by the vector field (\ref{Xvectorfield}), i.e., $MX$ has zero divergence, then
\begin{equation}\label{meqn}
\frac{d}{dt} \ln M + \frac{\partial X_i}{\partial x_i} = 0,
\end{equation} where $X_i = X_i(x_1,x_2,\cdots,x_m)$. 
\end{cor}

\textit{Proof --} Consider the dynamical vector field $X =  X_i \frac{\partial}{\partial x_i}$, where $X_i$ are suitable functions of the coordinates $x_i$ with $i=1,2,\cdots,m$. In local coordinates, the condition ${\rm div} (MX) = 0$ is equivalent to
\begin{equation}
 \frac{\partial}{\partial x_i} (M X_i) = 0,
\end{equation} and which gives (\ref{meqn}) or $X(M) + M ({\rm div} X) = 0$. 

\begin{cor}
The quantity $M \Omega$ is an invariant volume-form under the flow of $X$. 
\end{cor}

\textit{Proof --} Since $\pounds_{MX} \Omega = 0$, from Cartan's formula for the Lie derivative\footnote{Here, one makes use of Cartan's `magic' formula for the Lie derivative:
\begin{equation}\label{magic}
\pounds_X \alpha = d(\iota_X \alpha) + \iota_X (d\alpha),
\end{equation} where $\alpha$ is a differential form and $X$ is a vector field.}, we have $d(\iota_{MX} \Omega) = \pounds_X (M \Omega) = 0$. 

\subsection{Hamiltonian dynamics on symplectic manifolds}\label{symplecticsec}
Let us begin with a definition.

\begin{axiom}
A symplectic manifold is defined to be the pair $(\mathcal{M}_s,\omega)$, where $\mathcal{M}_s$ is a smooth manifold of (real) dimension $2n$ and $\omega$ is a two-form that is both closed and non-degenerate, i.e., 
\begin{equation}
d\omega = 0, \quad \quad \omega^n \neq 0. 
\end{equation}
\end{axiom}

The reader is referred to \cite{arnold,CM1,Geiges} for more details on symplectic manifolds. Let us recall below some useful facts which will be used in our subsequent analysis.

\subsubsection{Hamiltonian dynamics}
The non-degeneracy of $\omega$ allows one to define a vector-bundle isomorphism which induces a $C^\infty(\mathcal{M}_s,\mathbb{R})$-linear isomorphism between the set of vector fields and one-forms on $\mathcal{M}_s$. In particular, the Hamiltonian vector field $X_H$ is defined by
\begin{equation}\label{dHsymplecticdef}
\iota_{X_H}\omega = dH,
\end{equation} where $H \in C^\infty(\mathcal{M}_s,\mathbb{R})$. Darboux theorem asserts that near a point, one can find a local system of (Darboux) coordinates $(q^i,p_i)$ with $i \in \{1,2,\cdots,n\}$ such that 
\begin{equation}\label{omegadarboux}
\omega = dq^i \wedge dp_i.
\end{equation} Thus, the condition (\ref{dHsymplecticdef}) implies that 
\begin{equation}\label{XHlocal}
X_H = \frac{\partial H}{\partial p_i} \frac{\partial}{\partial q^i} - \frac{\partial H}{\partial q^i} \frac{\partial}{\partial p_i},
\end{equation} thereby indicating that one can recover the familiar Hamilton's equations as
\begin{equation}
\dot{q}^i = X_H(q^i) = \frac{\partial H}{\partial p_i}, \quad \quad \dot{p}_i = X_H(p_i) = -\frac{\partial H}{\partial q^i}. 
\end{equation}
Thus, the integral curves of the vector field $X_H$ satisfy the Hamilton's equations. An interesting consequence of (\ref{dHsymplecticdef}) is that by using the Cartan's formula (\ref{magic}), one gets
\begin{equation}
\pounds_{X_H} \omega = d(\iota_{X_H} \omega) + \iota_{X_H} d\omega = 0,
\end{equation} where the first term vanishes upon using (\ref{dHsymplecticdef}) because $d^2 = 0$ while the second term vanishes because $\omega$ is closed (by definition). The above-mentioned result implies that $\omega$ (hence the volume-form $\omega^n$) is conserved under the flow of $X_H$; this result is known as Liouville's theorem. 

\vspace{2mm}

Thus, a (conservative) Hamiltonian system may be formally defined as follows:

\begin{axiom}
A Hamiltonian system is the triple $(\mathcal{M}_s, \omega, H)$, where $(\mathcal{M}_s,\omega)$ is a symplectic manifold and $H \in C^\infty(\mathcal{M}_s, \mathbb{R})$. The corresponding dynamics is described by a Hamiltonian vector field which is determined by the condition $\iota_{{X}_H} \omega = dH$. 
\end{axiom}

\subsubsection{Conformal Hamiltonian dynamics}
The dynamics described in the preceding discussion is conservative in the sense that it not only conserves the Hamiltonian function as $X_H (H) = 0$, but it also preserves the phase-space volume. A simple step towards describing dissipative dynamics is to consider vector fields ${X}^\gamma_H$ that do not conserve the phase-space volume but satisfy \cite{conf,conf1}
\begin{equation}\label{conformalvolumedefinition1}
\pounds_{{X}^\gamma_H} \omega = - \gamma \omega,
\end{equation} where $\gamma$ is a real constant (the case $\gamma = 0$ corresponds to the previously-discussed case of a Hamiltonian vector field). Recall that in classical mechanics, the phase space $\mathcal{M}_s$ of a Hamiltonian system is the cotangent bundle of the configuration space $Q$, i.e., $\mathcal{M}_s = T^*Q$; it is naturally equipped with a tautological one-form $\theta$ which gives the symplectic two-form as $\omega = d\theta$. It may be noted that not all symplectic manifolds admit a symplectic two-form which is exact (although it is closed by definition); for closed symplectic manifolds (compact but without boundary) such as the two-sphere, the two-form $\omega$  is not exact. However, for our purposes from a mechanics viewpoint, it is useful to view the phase space as being a cotangent bundle in which case the symplectic two-form is exact\footnote{Note that Darboux theorem asserts that all symplectic manifolds of the same dimension (say, $2n$) are locally isomorphic to $T^* \mathbb{R}^n$. Thus, the symplectic two-form on the two-sphere can also be written as a derivative of a one-form but only locally -- recall that the two-sphere cannot be covered by a single chart.}; such symplectic manifolds are called exact symplectic manifolds. 

\vspace{2mm}

Now, for the phase space which is a cotangent bundle with tautological one-form $\theta$, we may define the vector field $X^\gamma_H$ alternatively as
\begin{equation}
\iota_{{X}^\gamma_H} \omega = dH - \gamma \theta,
\end{equation} which, upon plugging into Cartan's formula (\ref{magic}) gives (\ref{conformalvolumedefinition1}). In the Darboux coordinates (also called canonical coordinates), one has the standard expression of the tautological one-form which reads $\theta= -p_i dq^i$ and which is consistent with (\ref{omegadarboux}). This means the vector field ${X}^\gamma_H$ takes the following appearance:
\begin{equation}\label{confvecfield}
X^\gamma_H = \frac{\partial H}{\partial p_i} \frac{\partial}{\partial q^i} - \bigg(\frac{\partial H}{\partial q^i} + \gamma p_i\bigg) \frac{\partial}{\partial p_i},
\end{equation} and which means the integral curves of $X^\gamma_H$ are solutions of the equations
\begin{equation}\label{confeom}
\dot{q}^i = X^\gamma_H(q^i) = \frac{\partial H}{\partial p_i}, \quad \quad \dot{p}_i = X^\gamma_H(p_i) = -\frac{\partial H}{\partial q^i} - \gamma p_i.
\end{equation} For a mechanical system where $p_i$ are the momenta, one now finds a dissipation term with constant damping factor $\gamma$. This is known as conformal Hamiltonian dynamics \cite{conf}. Note that the conformal vector field (\ref{confvecfield}) may be expressed as 
\begin{equation}\label{conformalfromliouville}
X^\gamma_H = X_H - \gamma \Delta,
\end{equation} where $X_H$ is the standard (conservative) Hamiltonian vector field (\ref{XHlocal}) and $\Delta = p_i \frac{\partial}{\partial p_i}$ is the Liouville vector field. Viewing the phase space as a cotangent bundle $\pi: T^*Q \rightarrow Q$ on which $p_i$ are the induced fiber coordinates, i.e., $\pi: (q^i,p_i) \rightarrow q^i$, the vector field $\Delta$ generates dilatations along the fiber at each point. Here, the Darboux coordinates $(q^i,p_i)$ on the cotangent bundle $T^*Q$ are such that $q^i$ are the base-space coordinates (on $Q$) while $p_i$ are the fiber coordinates such that $\theta = -p_i dq^i$. Notice that one has the following relationship between the Liouville vector field and the tautological one-form on a cotangent bundle: 
\begin{equation}\label{Liovilledefinition}
\iota_\Delta \omega = \theta, \quad \quad \omega = d\theta,
\end{equation} and from Cartan's formula, this implies $\pounds_\Delta \omega = \omega$. With this background, let us furnish a formal definition of a conformal Hamiltonian system. 

\begin{axiom}\label{axiomconformal}
Consider an exact symplectic manifold $(\mathcal{M}_s,\omega)$ which is a cotangent bundle with tautological one-form $\theta$, i.e., $\omega = d\theta$. Then, a conformal Hamiltonian system is the quadruple $(\mathcal{M}_s, \omega, H, \gamma)$, where $H \in C^\infty(\mathcal{M}_s, \mathbb{R})$ and $\gamma \in \mathbb{R}$. The corresponding dynamics is described by a conformal vector field which is determined through the condition $\iota_{{X}^\gamma_H} \omega = dH - \gamma \theta$.
\end{axiom}

\textbf{Note:} As our motivation is to look at mechanical systems, we shall only be considering exact symplectic manifolds. Thus, we will not always specify the `exactness' of the symplectic two-form explicitly and will proceed with the understanding that all symplectic manifolds considered after this are exact and moreover, can be viewed as cotangent bundles. 

\subsection{Hamiltonian dynamics on contact manifolds}
Let us now formally define a contact manifold. 

\begin{axiom}\label{axiomcontactdef}
A contact manifold is a pair $(\mathcal{M}_c,\eta)$, where $\mathcal{M}_c$ is a smooth manifold of (real) dimension $2n+1$ and $\eta$ is a one-form satisfying
\begin{equation}\label{nonintegrability}
  \eta \wedge (d\eta)^n \neq 0.
\end{equation} 
\end{axiom}

\noindent
\textbf{Remarks:} Here, $\eta \wedge (d\eta)^n$ is the considered volume-form on $\mathcal{M}_c$. In the context of Frobenius integrability, the condition (\ref{nonintegrability}) means that the hyperplane distribution defined as ${\rm ker}(\eta)$ is maximally non-integrable\footnote{A more rigorous definition of a contact manifold relies on the existence of a hyperplane distribution which is maximally non-integrable. Such a distribution can be locally expressed as the kernel of a one-form satisfying the condition (\ref{nonintegrability}). If one can always write the associated hyperplane distribution as ${\rm ker}(\eta)$ (not just locally), then the contact manifold is called an exact contact manifold. In this paper, we will be considering only exact contact manifolds without further stating it explicitly.}. The reader is referred to \cite{arnold,Geiges,CM2} for more details.

\vspace{2mm}

On a contact manifold $(\mathcal{M}_c,\eta)$, there exists a vector field $\xi$ known as the Reeb vector field which is determined uniquely by the conditions
\begin{equation}\label{xidef}
  \iota_\xi \eta = 1, \quad \quad \iota_\xi d\eta = 0.
\end{equation} 
Darboux theorem asserts that if the condition (\ref{nonintegrability}) is satisfied, then it is possible to define local (Darboux) coordinates $(s,q^i,p_i)$ with $i \in \{1,2,\cdots,n\}$ near a point such that
\begin{equation}
  \eta = ds - p_i dq^i, \quad \quad \xi = \frac{\partial}{\partial s}.
\end{equation}

For a function $h \in C^\infty(\mathcal{M}_c,\mathbb{R})$, there is an associated vector field $X_h$ defined by the following combined conditions:
\begin{equation}\label{contactvecdef}
  \iota_{X_h} \eta = -h, \quad \quad \iota_{X_h} d\eta = dh - \xi(h) \eta.
\end{equation} The vector field $X_h$ is known as the contact vector field associated with the function $h$ and in local (Darboux) coordinates, it has the following expression:
\begin{equation}
  X_h = \bigg(p_i \frac{\partial h}{\partial p_i} - h \bigg) \frac{\partial }{\partial s} + \bigg(\frac{\partial h}{\partial p_i}\bigg)\frac{\partial }{\partial q^i} - \bigg( \frac{\partial h}{\partial q^i} + p_i \frac{\partial h}{\partial s} \bigg) \frac{\partial}{\partial p_i},
\end{equation} such that for any function $F \in C^\infty(\mathcal{M}_c,\mathbb{R})$, one has $X_h(F) = \frac{dF}{dt}$. The corresponding equations of motion are
\begin{equation}
\dot{s} = X_h (s) = p_i \frac{\partial h}{\partial p_i} - h, \quad \quad \dot{q}^i = X_h(q^i) = \frac{\partial h}{\partial p_i}, \quad \quad \dot{p}_i = X_h(p_i) = - \frac{\partial h}{\partial q^i} - p_i \frac{\partial h}{\partial s}. 
\end{equation}
Thus, given some `contact Hamiltonian' function $h \in C^\infty(\mathcal{M}_c,\mathbb{R})$, the relations (\ref{contactvecdef}) define a map $ h \mapsto X_h$ by which one can associate with it a contact vector field which yields certain equations of motion. We shall define a contact Hamiltonian system as follows: 

\begin{axiom}
A contact Hamiltonian system is the triple $(\mathcal{M}_c, \eta, h)$, where $(\mathcal{M}_c,\eta)$ is a contact manifold and $h \in C^\infty(\mathcal{M}_c,\mathbb{R})$. The dynamics is described by a contact vector field $X_h$ which is determined by the conditions $ \iota_{X_h} \eta = -h$ and $ \iota_{X_h} d\eta = dh - \xi(h) \eta$.
\end{axiom}

Below, let us point out some basic properties of a contact vector field. Clearly, $X_h$ does not conserve $h$ along its flow, i.e.,
\begin{equation}\label{dynh}
  X_h (h) = - h \xi(h) \neq 0,
\end{equation}
and which may also be seen without referring to the local (Darboux) coordinates just by contracting the second amongst equations (\ref{contactvecdef}) with $X_h$ and then using the first one. Furthermore, the flow of $X_h$ does not preserve the volume-form $\eta \wedge (d\eta)^n$ because
\begin{equation}
\pounds_{X_h} (\eta \wedge (d\eta)^n) = {\rm div}X_h  (\eta \wedge (d\eta)^n),
\end{equation} where the divergence of $X_h$ is found to be 
\begin{equation}\label{divcontactgen}
  {\rm div}X_h = -(n+1) \xi(h).
\end{equation} It is easy to see that if $\xi(h) = 0$, then the dynamics is conservative and the flow of $X_h$ also preserves the volume-form $\eta \wedge (d\eta)^n$. Moreover, note from (\ref{dynh}) that $h$ is conserved on the level set $\{h^{-1}(0)\}$, a feature that is exploited for the description of thermodynamic processes in the contact-geometric description of thermodynamics \cite{old2,new4,new6,new7,new11}. 

\subsection{Constant-damping systems}
A linearly-damped mechanical system with constant damping strength and one coordinate variable assumes the following second-order equation:
\begin{equation}\label{dampedeqgeneral}
m\ddot{q} + m \gamma \dot{q} + V'(q) = 0, \quad\quad \gamma > 0, \quad V(q) \in C^\infty(\mathbb{R}, \mathbb{R}),
\end{equation} where the prime denotes derivation with respect to $q$. The dynamics can be described by either a conformal vector field on a two-dimensional phase space $T^*\mathbb{R}$ endowed with a symplectic structure or a contact vector field on a three-dimensional (enlarged) phase space $T^*\mathbb{R} \times \mathbb{R}$ endowed with a contact structure. Let us consider the two cases one after the other. 

\subsubsection{Description via conformal vector field}\label{conformalexamplesec}
Consider a two-dimensional symplectic manifold $(\mathcal{M}_s,\omega)$ which is locally equivalent to $T^*\mathbb{R}$. Let $(q,p)$ be the Darboux coordinates near a point, i.e., one can write $\omega = dq \wedge dp$. Consider a Hamiltonian function $H \in C^\infty(\mathcal{M}_s,\mathbb{R})$ of the form
\begin{equation}\label{Hstandard}
H(q,p) = \frac{p^2}{2m} + V(q),
\end{equation} and a real constant $\gamma > 0$. From (\ref{confvecfield}), the corresponding conformal vector field takes the following appearance in the Darboux coordinates:
\begin{equation}
X^\gamma_H = \bigg( \frac{p}{m} \bigg) \frac{\partial}{\partial q} - \big(V'(q) + \gamma p  \big) \frac{\partial}{\partial p}.
\end{equation}
The equations of motion turn out to be
\begin{equation}
\dot{q} = \frac{p}{m}, \quad \quad \dot{p} = - V'(q) - \gamma p,
\end{equation} and the two may be combined to give (\ref{dampedeqgeneral}). Notice that $\omega = dq \wedge dp$ is not conserved under the dissipative dynamics described above. 

\subsubsection{Description via contact vector field}\label{secdampedcontact}
Consider a three-dimensional contact manifold $(\mathcal{M}_c,\eta)$ which is locally equivalent to $T^*\mathbb{R} \times \mathbb{R}$. In Darboux coordinates $(s,q,p)$, one has $\eta = ds - p dq$. Let us take a function $h \in C^\infty(\mathcal{M}_c,\mathbb{R})$ which reads
\begin{equation}
h(q,p,s) = \frac{p^2}{2m} + V(q) + \gamma s, \quad \quad \gamma > 0. 
\end{equation} The corresponding contact vector field reads as
\begin{equation}
X_h = \bigg(\frac{p^2}{2m} - V(q) - \gamma s \bigg) \frac{\partial }{\partial s} + \bigg(\frac{p}{m}\bigg) \frac{\partial }{\partial q} - \big( V'(q) + \gamma p \big) \frac{\partial}{\partial p}.
\end{equation}
Thus, the equations of motion are obtained as $\dot{s} = X_h(s)$, $\dot{q} = X_h(q)$, and $\dot{p} = X_h(p)$, which give
\begin{equation}
\dot{s} = \frac{p^2}{2m} - V(q) - \gamma s, \quad \quad \dot{q} = \frac{p}{m}, \quad \quad \dot{p} = -V'(q) - \gamma p.
\end{equation}
The relation $\dot{q} = p/m$ implies that $p$ is the linear momentum if one interprets $q$ as a mechanical (linear) coordinate. Combining this with the equation of motion for $p$ implies (\ref{dampedeqgeneral}) which represents the dynamics of a particle moving in a potential but under the influence of linear (constant) damping. If $h$ is independent of $s$, i.e., the Hamiltonian is of the standard type (\ref{Hstandard}), we have conservation of $h$ and the preservation of the phase-space volume $\eta \wedge d\eta = ds \wedge dq \wedge dp$. 

\section{Invariant measures from Jacobi last multiplier}\label{invariantsec1}
In this section, we will compute Jacobi last multipliers leading to the computation of invariant measures on the phase space for conformal Hamiltonian systems as well as for contact Hamiltonian systems. Let us take the two cases one after the other. 

\subsection{Conformal Hamiltonian dynamics}
Consider a conformal Hamiltonian system $(\mathcal{M}_s,\omega,H, \gamma)$. The corresponding conformal vector field $X_H^\gamma$ is given by (\ref{conformalfromliouville}), implying that 
\begin{equation}\label{dotHconformalliouville}
X^\gamma_H(H) = - \gamma \Delta(H),
\end{equation} and that ${\rm div} X^{\gamma}_H = - \gamma n$. From (\ref{meqn}), a Jacobi last multiplier should satisfy the following equation:
\begin{equation}\label{dMdtconformal}
X^\gamma_H(\ln M) = \gamma n.
\end{equation}
\begin{theorem}
Consider a conformal Hamiltonian system $(\mathcal{M}_s,\omega,H, \gamma)$ with ${\rm dim~} \mathcal{M}_s = 2n$. In the region of the phase space where $\Delta(H) \neq 0$, there is a Jacobi last multiplier given by
\begin{equation}\label{Mconformagenexp}
M = \exp \bigg( - n \int [\Delta(H)]^{-1} dH \bigg),
\end{equation} where $\Delta$ is the Liouville vector field.
\end{theorem}

\textit{Proof --} Consider the equation (\ref{dMdtconformal}). Combining this with (\ref{dotHconformalliouville}) and eliminating $\gamma$, one finds that
\begin{equation}\label{dMdtconformal1}
\frac{d}{dt} \ln M = - n \frac{dH}{dt} [\Delta(H)]^{-1},
\end{equation} or equivalently, 
\begin{equation}
d(\ln M) = - n [\Delta(H)]^{-1} dH.
\end{equation}
This immediately gives the result (\ref{Mconformagenexp}) upon integrating both sides. 

\begin{cor}
Corresponding to a conformal Hamiltonian system $(\mathcal{M}_s,\omega,H, \gamma)$, the phase-space measure in the region $\Delta(H) \neq 0$ which is invariant to the flow of the corresponding conformal vector field $X^\gamma_H$ is
\begin{equation}\label{invariantmeasureconformal}
\Omega|_{\Delta(H) \neq 0} =  \exp \bigg( - n \int [\Delta(H)]^{-1} dH \bigg) \omega^n,
\end{equation} where $\omega^n$ is a non-vanishing volume-form due to the non-degeneracy of the symplectic two-form.
\end{cor}
As a simple example, consider the linearly-damped dynamics of a free particle in spatial-dimension one $(n=1)$, i.e., the Hamiltonian is the same as (\ref{Hstandard}) but with $V(q) = 0$. Correspondingly, it is not hard to see that $\Delta(H) = 2 H$ (because the Hamiltonian is quadratic in the momentum) which means (\ref{invariantmeasureconformal}) suggests the following invariant measure on the two-dimensional phase space:
\begin{equation}\label{freeparticleconformalresult}
\Omega|_{H \neq 0} =  \exp \bigg( - \frac{1}{2} \int \frac{dH}{H} \bigg) \omega = \frac{\omega}{H^{1/2}}. 
\end{equation}

\noindent
\textbf{Remarks:} It turns out that if $V'(q) \neq 0$, then the integrand of (\ref{invariantmeasureconformal}) is not an exact differential and consequently, the expression (\ref{invariantmeasureconformal}) is of limited utility except for the free particle. However, for constant-damping systems in a non-trivial potential, invariant measure(s) can be found by considering either a contact-geometric description [Sec (\ref{secnewlineardampcontact}) below] or via a description in terms of generalized conformal vector fields as we will show in Sec. (\ref{constantdampingmeasuresec}); in the latter approach, the potential function must be carefully chosen so as to satisfy the Chiellini integrability condition (\ref{CC0}).

\subsection{Contact Hamiltonian systems}
In this section, let us discuss the role that Jacobi last multipliers play in the context of contact Hamiltonian systems in describing invariant phase-space measures. For a generic contact Hamiltonian system, i.e., the triple $(\mathcal{M}_c,\eta,h)$, equations (\ref{meqn}) and (\ref{divcontactgen}) imply that 
\begin{equation}\label{Mcontacteqn}
X_h( \ln M )= (n+1) \xi(h).
\end{equation}
Let us consider two distinct cases below. In particular, in Sec. (\ref{Mhnonzerosec1}), we will re-derive the `canonical invariant measure' as suited for contact Hamiltonian systems (presented originally in \cite{brav1}) using the framework of last multipliers. 

\subsubsection{Level set $\{h^{-1}(0)\}$}
In the region of the (contact) phase space where $h = 0$, (\ref{dynh}) implies that the flow of the contact vector field $X_h$ should be confined to within that region as $h$ is conserved when $h = 0$. In other words, the level set $\{h^{-1}(0)\}$ is invariant under the contact Hamiltonian dynamics. Formally, a Jacobi last multiplier is the solution of equation (\ref{Mcontacteqn}), giving $M = \exp \big(\int (n+1) \xi(h) dt \big)$. However, owing to the fact that the level set $\{h^{-1}(0)\}$ is a lower-dimensional subspace of the contact manifold $(\mathcal{M}_c,\eta)$, i.e., $\phi: \{h^{-1}(0)\} \hookrightarrow \mathcal{M}_c$, where $\phi$ is the inclusion map, we have $\phi^*(\eta \wedge (d\eta)^n) = 0$. In other words, $\eta \wedge (d\eta)^n$ is not a volume-form on the level set $\{h^{-1}(0)\}$. As a result, in what follows, we will restrict our attention to the region of the contact phase space given by $\mathcal{M}_c \setminus \{h^{-1}(0)\}$.

\subsubsection{Region $\mathcal{M}_c \setminus \{h^{-1}(0)\}$}\label{Mhnonzerosec1}
Let us discuss the situation outside the level set $\{h^{-1}(0)\}$, i.e., the region $\mathcal{M}_c \setminus \{h^{-1}(0)\}$ of the contact phase space. One finds the following result:
\begin{theorem}
Consider a contact Hamiltonian system $(\mathcal{M}_c,\eta,h)$ where ${\rm dim~} \mathcal{M}_c = 2n+1$. In the region $\mathcal{M}_c \setminus \{h^{-1}(0)\}$, there is a Jacobi last multiplier given by
\begin{equation}\label{Mhzero}
M = \frac{1}{h^{n+1}}.
\end{equation}
\end{theorem}

\textit{Proof --} The formal solution of the differential equation (\ref{Mcontacteqn}) is given by 
\begin{equation}
M = \exp \bigg(\int (n+1) \xi(h) dt \bigg).
\end{equation}
Substituting (\ref{dynh}) with $\frac{dh}{dt} = X_h(h)$, we find that 
\begin{equation}
M = \exp \bigg( -\int(n+1) h^{-1} dh \bigg),
\end{equation} and this directly gives the result (\ref{Mhzero}).  

\begin{cor}
In the region $\mathcal{M}_c \setminus \{h^{-1}(0)\}$, the following is an invariant measure:
\begin{equation}\label{resultnonzerohmeasure}
\Omega\big|_{h \neq 0} =  \frac{\eta \wedge (d\eta)^n}{h^{n+1}},
\end{equation} where $\eta \wedge (d\eta)^n$ is a non-vanishing volume-form by definition (\ref{nonintegrability}).
\end{cor}

\subsubsection{Constant-damping systems}\label{secnewlineardampcontact}
For a linearly-damped system with constant damping as treated in Sec. (\ref{secdampedcontact}), it is easy to apply the result (\ref{resultnonzerohmeasure}). On the three-dimensional phase space with $\eta \wedge d\eta = ds \wedge dq \wedge dp$, the invariant phase-space measure reads
\begin{equation}\label{resultnonzerohmeasurenew}
\Omega|_{h\neq 0} =  \bigg[\frac{p^2}{2m} + V(q) + \gamma s\bigg]^{-2}(ds \wedge dq \wedge dp), \quad \quad h = \frac{p^2}{2m} + V(q) + \gamma s \neq 0.
\end{equation}

\section{Li\'enard systems and generalized conformal vector fields}\label{invariantsec2}
We will now describe Li\'enard-type systems and formulate an appropriate geometric setting for their description. This will allow us to present invariant phase-space measures in explicit form when a certain integrability condition is obeyed. 

\subsection{Generalized conformal vector fields}
We can now present the notion of a `generalized' conformal vector field which was also briefly discussed in \cite{CM5}. 

\begin{axiom}\label{axiomconformal0}
Consider a configuration space $Q$ and a corresponding phase space $T^*Q$ endowed with a symplectic two-form $\omega$. A generalized conformal vector field $Y$ is a vector field such
that there exists a function $K$ on $Q$ such that
\begin{equation}\label{congendef1}
\pounds_Y\omega = - K \omega,
\end{equation}
and then there exists a (locally defined) function $H$ in $T^*Q$ such that $\iota_Y \omega = dH - K \theta$ if $dK \wedge \theta = 0$. Such a vector field $Y$ will be denoted by $X^K_H$.
\end{axiom}

\begin{cor}\label{cor41}
Consider the cotangent bundle $T^*\mathbb{R}$ which is naturally equipped with the tautological one-form $\theta$ leading to the symplectic two-form $\omega = d\theta$. If $X^K_H$ be a generalized conformal vector field and if the function $K$ defined on $\mathbb{R}$ is differentiable, then the condition $\pounds_{X^K_H} \omega = -K \omega$ is compatible with $\iota_{X^K_H} \omega = dH - K \theta$ because $dK \wedge \theta = 0$.
\end{cor}

\textit{Proof --} Consider the two-dimensional phase space $\pi: T^*\mathbb{R} \rightarrow \mathbb{R}$. If $q$ be a coordinate on $\mathbb{R}$ and $\pi:(q,p) \rightarrow q$ be the corresponding `induced' fiber coordinates on the cotangent bundle, then naturally, $\theta = - p dq$ (implying $\omega = d\theta = dq \wedge dp$) from the coordinate definition of the tautological one-form on $T^*\mathbb{R}$. Thus, given a differentiable function $K(q)$ on $\mathbb{R}$, one has $dK \wedge \theta = 0$. Consequently, from Cartan's formula (\ref{magic}) for the Lie derivative, one can write $\pounds_{X^K_H} \omega = d(\iota_{X^K_H} \omega ) + \iota_{X^K_H} (d\omega)$, implying that $-K \omega = d(\iota_{X^K_H} \omega )$. The two sides of this equation agree if we put $\iota_{X^K_H} \omega = dH - K \theta$ because $dK \wedge \theta = 0$.

\vspace{2mm}

\noindent
Since our interest is in Li\'enard-type systems which admit one-dimensional configuration spaces, we will focus hereafter on phase spaces that are $T^*\mathbb{R}$ (or a subspace of it) for which the above-mentioned corollary holds. For some $H = H(q,p)$ and $K=K(q)$, the equations of motion turn out to be 
\begin{equation}\label{genconfonedeqn}
\dot{q} = X^K_H(q) =  \frac{\partial H}{\partial p}, \quad \quad \dot{p} = X^K_H(p) = - \frac{\partial H}{\partial q} - K(q) p.
\end{equation}
Interpreting $q$ as a mechanical coordinate with $p$ being the corresponding momentum, one finds that the second equation above describes dynamics with damping linear in the momentum but with a position-dependent damping strength. When $K$ is a constant, the dynamics described above reduces to that dictated by a conformal vector field (\ref{confvecfield}) discussed earlier. 

\vspace{2mm}

Using the Liouville vector field $\Delta$ on $T^*\mathbb{R}$ as defined in (\ref{Liovilledefinition}), one can express a generalized conformal vector field as
\begin{equation}\label{Genconcoord}
X^K_H = X_H - K \Delta, 
\end{equation} where $X_H$ is the (conservative) Hamiltonian vector field corresponding to the function $H \in C^\infty(T^*\mathbb{R}, \mathbb{R})$. Thus, it turns out that the Hamiltonian is not conserved, i.e., 
\begin{equation}
X^K_H (H) = - K \Delta (H), 
\end{equation} and from (\ref{congendef1}), the divergence of $X^K_H$ turns out to be ${\rm div} X^K_H = - K$, where $\omega$ is the chosen volume-form on $(\mathcal{M}_s = T^* \mathbb{R}, \omega)$ which is not invariant under the flow of $X^K_H$. 

\subsection{Geometric description of Li\'enard systems}
 Choosing a standard Hamiltonian as in (\ref{Hstandard}), we find that the generalized conformal vector field (\ref{Genconcoord}) takes the following appearance in Darboux coordinates:
 \begin{equation}
 X^K_H = \bigg(  \frac{p}{m} \bigg) \frac{\partial}{\partial q} - (V'(q) + K(q) p) \frac{\partial}{\partial p}. 
 \end{equation}
 Thus, the corresponding equations of motion turn out to be
\begin{equation}\label{Lienardconf1storder}
\dot{q} =  \frac{p}{m}, \quad \quad \dot{p} = - V'(q) - K(q) p.
\end{equation}
Combining the two equations above, we get
\begin{equation}
\ddot{q} + f(q) \dot{q} + g(q) = 0,
\end{equation} where $f(q) = K(q)$ and $g(q) = V'(q)/m$. Thus, we have found a geometric description of the Li\'enard system (\ref{Lienard1definition}) using generalized conformal vector fields which satisfy the condition (\ref{congendef1}). 

\subsection{Chiellini integrability condition and invariant measures}
We will prove the following result: 

\begin{theorem}
Consider a Li\'enard system as described by a generalized conformal Hamiltonian system $(\mathcal{M}_s, \omega, H, K)$, where $\mathcal{M}_s \approxeq T^* \mathbb{R}$ and $\theta = -p dq$ is the tautological one-form leading to the symplectic two-form $\omega = d\theta = dq \wedge dp$ on $\mathcal{M}_s$; the Hamiltonian function is given by (\ref{Hstandard}) and $K(q)$ is some suitable non-zero function of $q$. If the following condition is satisfied: 
\begin{equation}\label{CC}
\frac{d}{dq} \bigg( \frac{V'(q)}{m K(q)} \bigg) + l (l + 1) K(q) = 0,
\end{equation} then the following is an invariant measure on the phase space:
\begin{equation}\label{Lienardmeasureinvariant}
\Omega = \bigg(p - \frac{V'(q)}{lK(q)}\bigg)^{1/l} (dq \wedge dp),
\end{equation} where $l$ is determined through the condition (\ref{CC}). 
\end{theorem}

\textit{Proof --} For the system (\ref{Lienardconf1storder}), the divergence in local coordinates $(q,p)$ turns out to be
\begin{equation}
{\rm div} X^K_H = - K(q). 
\end{equation}
Thus, the equation (\ref{meqn}) for the last multiplier becomes
\begin{equation}
\frac{d}{dt}\ln M = K(q),
\end{equation} and which can be formally integrated to give
\begin{equation}\label{Mlienardoneformal}
M = \exp \bigg( \int K(q) dt \bigg).
\end{equation}
Let us define a new variable $u$ as \cite{JLMLienard}
\begin{equation}
m u = p - G(q), \quad \quad G(q) = \frac{V'(q)}{lK(q)},
\end{equation} where $l$ is determined from the condition (\ref{CC}). If $K(q)$ and $V(q)$ are such that (\ref{CC}) is satisfied, we then get $\dot{u} = l u K(q) $ along with $p = m u + G(q)$ which is equivalent to the Li\'enard system. However, now that one has $\dot{u} = l u K(q) $, we can write $K(q) dt = (lu)^{-1} du$ and substituting this into (\ref{Mlienardoneformal}) gives us $M = u^{1/l}$ or equivalently,
\begin{equation}\label{lienardM}
M(q,p) = \bigg(\frac{p}{m} - \frac{V'(q)}{mlK(q)}\bigg)^{1/l},
\end{equation} where notice that the last multiplier has been expressed as a function of the phase-space variables $(q,p)$. Since $m$ is a constant, the invariant phase-space measure associated with the system (\ref{Lienardconf1storder}) turns out to be (\ref{Lienardmeasureinvariant}). 

\vspace{2mm}

\noindent
\textbf{Remarks:} The expressions (\ref{Lienardmeasureinvariant}) and (\ref{lienardM}) are true only when the condition (\ref{CC}) is true; (\ref{CC}) is equivalent to (\ref{CC0}) which is termed the Chiellini integrability condition (see for example, \cite{CM3,CM5,JLMLienard,mitra24}). Notice that generally two values of $l$ are obtained from (\ref{CC}) and therefore the invariant measures are not unique.

\subsubsection{Constant-damping systems in quadratic and constant potentials}\label{constantdampingmeasuresec}
If one considers the case of constant damping as a special case of the Li\'enard system, then putting $f(q) = K(q) = \gamma > 0$ implies that the Chiellini integrability condition (\ref{CC}) demands
\begin{equation}
V''(q) + m \gamma^2 l (l+1) = 0.
\end{equation}
Let us take two cases one after the other.

\vspace{2mm}

\textbf{Quadratic potential:} Choosing $V(q) = \alpha_0 q^2$ where $\alpha_0$ is a constant, one gets the quadratic equation
\begin{equation}
l^2 + l + \frac{2 \alpha_0}{m \gamma^2} = 0. 
\end{equation}
This generally gives two values for $l$, whose reality imposes the condition $m \gamma^2 \geq 8 \alpha_0$. For simplicity, choosing $\alpha_0 = m \gamma^2/8$ so that there is only one value of $l$, namely, $l = -1/2$, the expression (\ref{lienardM}) gives
\begin{equation}
M(q,p) = \bigg[ p + \frac{m \gamma q}{2} \bigg]^{-2},
\end{equation} up to a constant factor. Using this, it is easy to check that $X^{K = \gamma}_H (\ln M) = \gamma$, which satisfies the equation (\ref{meqn}). Thus, the invariant measure (\ref{Lienardmeasureinvariant}) is
\begin{equation}
\Omega =  \bigg[ p + \frac{m \gamma q}{2} \bigg]^{-2} (dq \wedge dp),
\end{equation} subject to $p \neq - \frac{m \gamma q}{2}$. Notice that we cannot treat general potentials here as they will not satisfy the Chiellini integrability condition (\ref{CC}). 

\vspace{2mm}

\textbf{Constant potential:} If $V'(q) = 0$, the Chiellini integrability condition (\ref{CC}) implies $l = 0, -1$. Referring to equation (\ref{lienardM}), one must discard $l = 0$ and take $l = -1$, giving $M \sim 1/p $. Thus, the result (\ref{Lienardmeasureinvariant}) now corresponds to the result (\ref{freeparticleconformalresult}) obtained earlier for the free particle with constant damping. 

\section{Discussion}\label{summary}
In this paper, we have presented invariant phase-space measures for certain mechanical systems using the framework of Jacobi last multipliers. In particular, we analyzed conformal Hamiltonian dynamics as well as contact Hamiltonian dynamics. Another class of systems that we analyzed were systems of the Li\'enard class which may possess position-dependent damping. For such systems, we discussed a geometric description of the dynamics on cotangent bundles by presenting the notion of a generalized conformal vector field. This also allowed us to obtain analytical expressions for invariant phase-space measures associated with such systems when the Chiellini integrability condition is satisfied. A closely-related analysis may be carried out for a class of Levinson-Smith equations discussed recently in \cite{mitra24}. 

\vspace{2mm}

Let us end by interpreting the notion of last multipliers from the point of view of a `generalized' Liouville equation. In particular, given a dynamical system whose phase trajectories are the integral curves of a vector field $X$, the equation (\ref{meqn}) may be re-written as
\begin{equation}\label{liou}
X(\ln |M|) + {\rm div} X = 0,
\end{equation} where we have taken the modulus sign in writing $|M|$ as is necessary for interpreting it as a phase-space density. 
The above-mentioned equation (\ref{liou}) may be straightforwardly interpreted as a generalized Liouville equation (see also, \cite{hamel}) where $|M|$ can be viewed as the phase-space density. For conservative mechanical systems described by Hamiltonian dynamics on symplectic manifolds (where $X = X_H$; as defined in (\ref{dHsymplecticdef})), from $\pounds_{X_H} \omega^n = 0$, one has ${\rm div} X_H = 0$ which gives $X_H(\ln |M|) = 0 = X_H(|M|)$. This is the Liouville equation that one encounters in classical mechanics and statistical mechanics, wherein one identifies the phase-space density $\rho = |M|$. Explicitly, one has
\begin{equation}
\frac{d\rho}{dt} = 0,
\end{equation} which is foundational to the formulation of statistical mechanics. In the present study, we have considered systems that cannot be described from the point of view of Hamiltonian dynamics on symplectic manifolds and have presented explicit results for invariant phase-space measures $\{\Omega_{\rm Invariant} = M \Omega\}$ starting with volume-forms $\{\Omega\}$ that are not conserved under the dynamics, i.e., $\pounds_{X} \Omega \neq 0$. Different cases for $X$ as discussed in this work are conformal ($X^\gamma_H$), generalized conformal ($X^K_H$), and contact ($X_h$) vector fields. Thus, the present exposition can be understood as a preliminary step towards a formulation of statistical mechanics for dissipative systems (see for example, \cite{brav1,bravstat}).

\section*{Acknowledgements} The financial support received from the Ministry of Education (MoE), Government of India in the form of a Prime Minister's Research Fellowship (ID: 1200454) is gratefully acknowledged. The author also thanks the School of Physics, University
of Hyderabad for hospitality and local-travel support through IoE-UoH-IPDF (EH) scheme.


\begin{thebibliography}{99}


\bibitem{arnold} V. I. Arnold, {\it Mathematical Methods of Classical Mechanics}, 2nd ed., Springer (1989).

\bibitem{conf} R. McLachlan and M. Perlmutter, J. Geom. Phys. \textbf{39}, 276 (2001). 

\bibitem{conf1} J. F. Cari\~nena, F. Falceto, and M. F. Ra\~nada, J. Geom. Mech. \textbf{5}, 151 (2013). 

\bibitem{CM1} M. de Le\'on and C. Sard\'on, J. Phys. A: Math. Theor. \textbf{50}, 255205 (2017).
            
\bibitem{biology} P. Guha and A. Ghose-Choudhury, J. Geom. Phys. \textbf{134}, 195 (2018).

\bibitem{Geiges} H. Geiges, {\it An Introduction to Contact Topology}, Cambridge University Press (2008).

\bibitem{CM2} A. Bravetti, H. Cruz, and D. Tapias, Ann. Phys. (N.Y.) \textbf{376}, 17 (2017).

\bibitem{CM3} J. F. Cari\~nena and P. Guha, Int. J. Geom. Methods Mod. Phys. \textbf{16}, 1940001 (2019).

\bibitem{CM4} J. Gaset, X. Gr\`acia, M. C. Mu\~noz-Lecanda, X. Rivas, and N. Rom\'an-Roy, Int. J. Geom. Methods Mod. Phys. \textbf{17}, 2050090 (2020).

\bibitem{CM5} J. F. Cari\~nena and P. Guha, Int. J. Geom. Methods Mod. Phys. \textbf{21}, 2440005 (2024).

\bibitem{brav1} A. Bravetti and D. Tapias, J. Phys. A: Math. Theor. \textbf{48}, 245001 (2015).

\bibitem{brav2} A. Bravetti, M. de Le\'on, J. C. Marrero, and E. Padr\'on, J. Phys. A: Math. Theor. \textbf{53}, 455205 (2020).

\bibitem{bravstat} A. Bravetti and D. Tapias, Phys. Rev. E \textbf{93}, 022139 (2016).

\bibitem{chandra}V. K. Chandrasekar, M. Senthilvelan, and M. Lakshmanan, Phys. Rev. E \textbf{72}, 066203 (2005).

\bibitem{limit} L. Perko, {\it Differential Equations and Dynamical Systems}, 3rd ed., Springer (2001).

\bibitem{whittaker} E. T. Whittaker, {\it A Treatise on the Analytical Dynamics of Particles and Rigid Bodies}, Cambridge University Press (1988).

\bibitem{Akash} A. Sinha and A. Ghosh, Pramana \textbf{98}, 101 (2024).   

\bibitem{Leach1} M. C. Nucci and P. G. L. Leach, Phys. Scr. \textbf{78}, 065011 (2008).

\bibitem{Leach2} M. C. Nucci and P. G. L. Leach, J. Math. Phys. \textbf{49}, 073517 (2008).

\bibitem{Leach3} M. C. Nucci and P. G. L. Leach, J. Nonlinear Math. Phys. \textbf{16}, 431 (2009).

\bibitem{AGC} A. Ghose Choudhury, P. Guha, and B. Khanra, J. Math. Anal. Appl. \textbf{360}, 651 (2009).

\bibitem{JLMLienard} M. C. Nucci and K. M. Tamizhmani, J. Nonlinear Math. Phys. \textbf{17}, 167 (2010).

\bibitem{mitra24} S. Mitra, A. Ghose-Choudhury, S. Poddar, S. Garai, and P. Guha, Phys. Scr. {\bf 99}, 015237 (2024).

\bibitem{yan78} C. C. Yan, Am. J. Phys. {\bf 46}, 671 (1978).

\bibitem{PG3} P. Guha and A. Ghose Choudhury, Rev. Math. Phys. \textbf{25}, 1330009 (2013).
         
\bibitem{PG4} A. Ghose Choudhury and P. Guha, J. Phys. A: Math. Theor. \textbf{46}, 165202 (2013).
           
\bibitem{hamel} J. F. Cari\~nena and P. Santos, J. Phys. A: Math. Theor. \textbf{54}, 225203 (2021). 
           
\bibitem{carinenahojman} J. F. Cari\~nena and M. F. Ra\~nada, Int. J. Geom. Methods Mod. Phys. \textbf{18}, 2150166 (2021).
         
\bibitem{Nucci} M. C. Nucci, J. Nonlinear Math. Phys. \textbf{12}, 284 (2005). 
                 
\bibitem{JLMNew} G. Gonz\'alez Contreras and A. Yakhno, Symmetry \textbf{15}, 1416 (2023).
            
\bibitem{carinenaintegrability} J. F. Cari\~nena and J. Fern\'andez-N\'u\~nez, Symmetry \textbf{13}, 1413 (2021).

\bibitem{old2} R. Mruga\l{}a, Rep. Math. Phys. \textbf{33}, 149 (1993).
       
\bibitem{new4} A. Bravetti, C. S. Lopez-Monsalvo, and F. Nettel, Ann. Phys. (N.Y.) \textbf{361}, 377 (2015). 
                 
\bibitem{new6} A. Bravetti, Int. J. Geom. Methods Mod. Phys. \textbf{16}, 1940003 (2019).
        
\bibitem{new7} A. Ghosh and C. Bhamidipati, Phys. Rev. D \textbf{100}, 126020 (2019). 
                          
\bibitem{new11} A. Ghosh, Pramana \textbf{97}, 49 (2023).

\end{thebibliography}
\end{document}